\documentclass[doublecol]{epl2} 
\usepackage{amsmath,amsthm,amssymb}
\usepackage{color}

\title{Can we study the many-body localisation transition?}

\author{R.~K.~Panda \inst{1,2} \and A.~Scardicchio \inst{1,3} \and M.~Schulz \inst{1} \and S.~R.~Taylor \inst{1} \footnote{Corresponding author. Email: staylor@ictp.it} \and M.~ \v Znidari\v c \inst{4}}
\shortauthor{R.~K.~Panda \etal}

\institute{                    
  \inst{1} The Abdus Salam International Centre for Theoretical Physics - Strada Costiera 11, 34151 Trieste, Italy\\
  \inst{2} SISSA - Via Bonomea 265, 34136 Trieste, Italy\\
  \inst{3} INFN, Sezione di Trieste - Via Valerio 2, 34126 Trieste, Italy\\
  \inst{4} Physics Department, Faculty of Mathematics and Physics, University of Ljubljana - Jadranska 19, SI-1000 Ljubljana, Slovenia
}


\abstract{
We present a detailed analysis of the length- and timescales needed to approach the critical region of MBL from the delocalised phase, studying both eigenstates and the time evolution of an initial state. For the eigenstates we show that in the delocalised region there is a single length, which is a function of disorder strength, controlling the finite-size flow. Small systems look localised, and only for larger systems do resonances develop which restore ergodicity in the form of the eigenstate thermalisation hypothesis. For the transport properties, we study the time necessary to transport a single spin across a domain wall, showing how this grows quickly with increasing disorder, and compare it with the Heisenberg time. For a sufficiently large system the Heisenberg time is always larger than the transport time, but for a smaller system this is not necessarily the case. We conclude that the properties of the MBL transition cannot be explored using the system sizes or times available to current numerical and experimental studies. 
}

\begin{document}

\maketitle

\section{Introduction}
The question of whether or not a generic, isolated quantum system must necessarily reach thermal equilibrium, as described in the canon of Statistical Mechanics \cite{landau2013course,rigol2008thermalization,Goldstein2015}, has received a great deal of attention. This question is important both for our understanding of the fundamental laws of Nature (which are quantum mechanical at their core) and for applications in the growing field of Quantum Technologies. Counterexamples in various forms have been found as time-translation invariant systems with strong disorder (many-body localised systems \cite{Basko:2006hh,gornyi2005interacting,de2013ergodicity,huse2014phenomenology,nandkishore2015many,luitz2015many,abanin2017recent}), periodically driven systems (time crystals \cite{zhang2017observation,sacha2017time,Abanin2016Theory}), and even systems without disorder \cite{brenes2018many,Papic2015,schiulaz2014glass,pino2015metallic,yao2016quasi,nandkishore2017amany,Schulz2019Stark,vanNieuwenburg2019From}.  
Many-body localised systems, the focus of this letter, possess an extensive set of quasilocal conserved quantities (so-called local integrals of motion) \cite{serbyn2013local,imbrie2014many,ros2015integrals,imbrie2017local}, and thus exhibit no transport of conserved quantities such as particles or energy, but only of entanglement \cite{znidaric08}.

The ergodic phase of these systems, found at weak disorder strengths preceding the many-body localisation (MBL) transition, also exhibits fascinating phenomena, with numerical studies observing anomalous subdiffusive transport \cite{Reichman2014Absence,Varma2017Energy,AgarwalAnomalousDiffusion,Znidaric2016Diffusive,Schulz2018Energy,Mendoza2019Asymmetry,Schulz2019Phenomenology}, yet also volume-law entanglement entropy of eigenstates \cite{luitz2015many}. Despite the fact that there is compelling evidence for both the subdiffusive phase, dominated by an unusually slow (but complete) decay of excitations, and for the freezing of excitations for very strong disorder (at least for the small system sizes in numerical studies or small times observed in cold atoms experiments \cite{Bloch2015}), the properties of the phase transition separating the ergodic and MBL phases remain elusive \cite{Khemani2017Critical}. This has led to a fierce debate in the community \cite{vsuntajs2019quantum,Abanin2019Distinguishing,Sierant2019Thouless} on the nature and even the existence of an MBL transition.

For lack of a theory of the transition, all numerical studies based on exact diagonalisation (ED) \cite{pietracaprina2018shift} have adopted a single-parameter scaling to achieve a data collapse.
It is assumed that a single lengthscale $\xi \propto |W-W_c|^{-\nu}$, where $W$ is a measure of the strength of the disorder, diverges at the transition, and all other quantities (entanglement entropy, correlation functions, level statistics etc.) are functions of $L/\xi$, where $L$ is the system size.
The collapse thus obtained is usually quite good (see for example \cite{alet2015, Khemani2017Critical, Pietracaprina2017Entanglement, Mace2019Multifractal}) but the critical exponents contradict simple bounds set by general considerations \cite{chandran2015finite}: numerical works find $\nu\simeq 1$, while the arguments in Ref.~\cite{chandran2015finite} would imply $\nu\geq 2$.
A possible explanation of this contradiction is given by the strong-disorder renormalisation group (SDRG) picture of the MBL transition \cite{Altman2015,AltmanTheory2015,dumitrescu2017scaling,thiery2018many}, which can imply that the transition has the characteristics of a Kosterlitz-Thouless transition \cite{dumitrescu2019kosterlitz,goremykina2019analytically}. This transition is known to have particularly bad (logarithmic) corrections, which could explain how numerical methods are fooled into finding unphysical critical exponents.

In this letter we study the disordered spin-1/2 Heisenberg chain:
\begin{equation}
    H=J\sum_{n=1}^L \vec{s}_n\cdot\vec{s}_{n+1}+h_n s^z_n,
    \label{eq:Heisenberg}
\end{equation}
where $L$ is the system size, $J=1$, and $h_n \in [-W, W]$ are uniformly-distributed independent random numbers.
We argue that the physics of the MBL transition is not accessible by current numerical and experimental methods, and that even approaching the transition from the ergodic side is unfeasible with methods like ED. Depending on the true critical disorder, large system sizes are needed to make any claim about the physics close to the transition. Under reasonable assumptions, we suggest systems sizes $L \gtrsim 40$ are necessary.
We reach this conclusion on two independent fronts. First we analyse eigenstates and in particular the validity of the eigenstate thermalisation hypothesis (ETH), with a slightly different focus from previous works \cite{Luitz2016Anomalous,colmenarez2019statistics,Foini2019Eigenstate,Luitz2019Multifractality}. We show that for $W\leq 2.4$ (well within the subdiffusive region \cite{Znidaric2016Diffusive,Schulz2019Phenomenology}), if the system size is larger than a critical length $L_0$, then the ETH is fully recovered and the matrix elements of local operators have a Gaussian distribution. At $W=2.8$ we have $L_0\simeq 22$, which is the largest system size we can analyse, and $L_0(W)$ is very much linear in $W$ up to this disorder strength. We observe no signature of the upward curvature preceding the putative MBL transition at $W=W_c\sim 4$, suggesting that the transition must be at significantly larger disorder strengths (this is corroborated by the second part of the letter, where a quantity analogous to $L_0(W)$ shows curvature only for $W\gtrsim 2.5$).
On the second front, we quantitatively study the slow transport on the ergodic side. We argue that the dynamics could be so slow that the ergodic time becomes larger than the Heisenberg time for small systems and large disorder, and one cannot make reliable statements about the true behaviour in the thermodynamic limit (TDL) because Heisenberg recurrences can be mistaken for localisation. The Heisenberg time has been compared to times over which chaos is developed in Ref.~\cite{vsuntajs2019quantum} for a range of $L,W$, and it is conjectured that for any $L$ there is always a disorder value $W$ for which this is true. Our main observation is rather different: one cannot extract the location (or even the existence) of the transition from small ED studies.

\section{Lengthscales near the MBL transition}
In an isolated quantum system, the behaviour of the expectation value of an operator, $\langle \hat{A} \rangle (t) = \langle \psi(t) | \hat{A} | \psi(t) \rangle$, is most easily understood in the eigenbasis of the system's Hamiltonian (eigenstates $| \alpha \rangle$ with eigenenergies $E_{\alpha}$): the diagonal elements of $\hat{A}$ govern the long-time average value, while the off-diagonal elements are associated with the relaxation to this value.
The ETH ansatz describes the statistical properties of matrix elements in this basis.
If the matrix obeys the ETH ansatz, then the expectation value will thermalise under unitary dynamics.
The ansatz has the form:
 \begin{equation}
     \langle \alpha | \hat{A} | \beta \rangle = \overline{A}(E) \delta_{\alpha,\beta} + {\rm e}^{-s(E) L / 2} f(E,\omega) R_{\alpha,\beta},
 \end{equation}
where $E = (E_{\alpha} + E_{\beta})/2$, $\omega=E_{\beta}-E_{\alpha}$, $s(E)$ is the microcanonical entropy density, $L$ is the system volume, $\overline{A}(E)$ and $f(E,\omega)$ are smooth functions of their arguments, and $R_{\alpha,\beta}$ is a random variable with zero mean and unit variance.
Consider the connected correlation function of $\hat{A}$ in an eigenstate:
\begin{eqnarray}
\label{eq:ccETH}
    \langle \alpha | \hat{A}(t)\hat{A}(0) | \alpha \rangle_c&=&\sum_{\beta\neq\alpha}{\rm e}^{-i(E_\alpha-E_\beta)t}|\langle \alpha | \hat{A} | \beta \rangle|^2 \nonumber\\
    &=&\int_{-\infty}^{\infty} d\omega {\rm e}^{i\omega t}f(E,\omega)^2,
\end{eqnarray}
where we have replaced $R_{\alpha,\beta}^2$ with its mean, 1, and $f^2$ is then interpreted as the structure factor associated with the correlation function (we assume $E_\alpha$ is an infinite temperature state). Under general assumptions about the equilibrium state, the connected correlation function must decay to zero as $t\to\infty$ and the mean value $\overline{A}(E)$ must be equal to the microcanonical expectation value of $\hat{A}$ at energy $E$.

The distribution of $R_{\alpha,\beta}$ is assumed to be Gaussian \cite{Deutsch1991,Srednicki1994,Berry1977Regular}, and recent studies have linked subdiffusive transport of conserved quantities to non-Gaussian (\emph{i.e.}\ power-law) distributions of the operator's off-diagonal matrix elements. However, these studies were limited to values of $\omega\propto \Delta E,$ the mean level spacing, which is exponentially small in $L$ \cite{Luitz2016Anomalous}.
Correlation functions of local operators in the ETH have a well defined limit as $L\to\infty$; in particular, for $t=O(1)$ as $L\to\infty$, the contributions to the integral \eqref{eq:ccETH} from an interval of $\omega\sim O(L^{-a})$ are negligible.
On the other hand, the matrix elements at very small $\omega$ determine transport properties (or large time $t\sim L^a$ behaviour) as can be seen in the limit of DC conductivity (\emph{i.e.}\ when $\omega\to 0$) in the Kubo formula. An anomalous small-$\omega$ behaviour is indeed linked to a sub-ohmic conductivity $\sigma\propto L^{-\gamma}$, with $\gamma>1$, signalling subdiffusive transport. To the best of our knowledge the ETH and subdffusive transport are compatible, and a system whose local correlation functions equilibrate can indeed present slow dynamics at global, $O(L^{\gamma+1})$-time scales.

We find that if one focuses on matrix elements at finite energy differences, say $\omega=1$ (no significant differences are observed for $\omega=0.25,0.5,1.5$, as shown in the Supplemental Material; the data can be found at \cite{EDData}), if the system size $L$ is sufficiently large, the Gaussian distribution is recovered, and therefore the power-law distribution must be confined to a vanishing interval of $\omega$.

We obtain $R_{\alpha,\beta}=\langle\alpha|s^z_{L/2}|\beta\rangle / \sigma (\langle\alpha|s^z_{L/2}|\beta\rangle)$ (where $\sigma (\langle\alpha|s^z_{L/2}|\beta\rangle)$ is the standard deviation of the matrix elements) by numerical ED on small systems ($L \leq 22$), using the shift-invert technique to target eigenstates at a chosen energy \cite{abhyankar2018petsc, petsc-efficient, SLEPc, Hernandez2003SLEPc, slepc-users-manual, pietracaprina2018shift, MUMPs_1, MUMPs_2}.
We work in the zero-magnetisation sector, which has a Hilbert space dimension of $\mathcal{D}_L = \genfrac(){0pt}{2}{L}{L/2}$, applying open boundary conditions. We calculate $R_{\alpha,\beta}$ between 50 states closest to two target energies from near the middle of the spectrum; for the first target we choose the average energy in the sector, $E_0 = \mathrm{Tr} (H) / \mathcal{D} = -1/4$, and the second target is at $E_0 + \omega$.

We quantify the shape of the distribution with the Binder cumulant:
\begin{equation}
    B = \frac{\left\langle R^4 \right\rangle}{3 \left\langle 
    R^2 \right\rangle^2},
\end{equation}
where the angled brackets denote an average. The closer $B$ is to 1, the better the Gaussian distribution is realised. 

For each disorder realisation, $B$ is found by averaging over the $50 \times 50$ off-diagonal matrix elements. Here we focus on the median value of $B$ (we find no significant differences in the behaviour of the average). We notice that the Binder cumulant is always larger than 1, or consistent with 1 within errors, so we define $C := B - 1$. 

\begin{figure}
    \centering
    \includegraphics[width=\columnwidth]{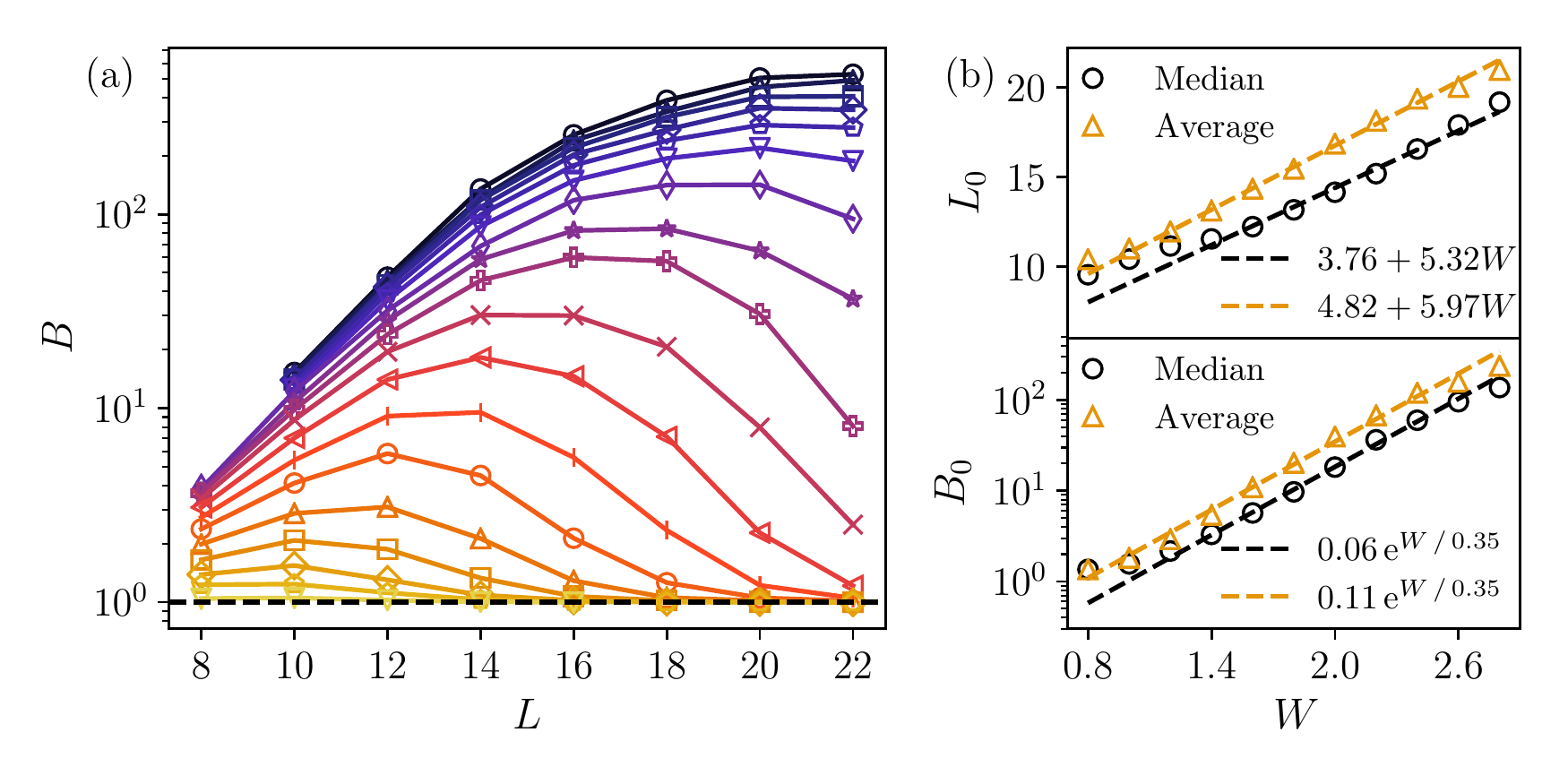}
    \caption{Behaviour of the Binder cumulant $B$. (a) Dependence of the median value of $B$ on system size $L$ for several disorder strengths $W$, which tends to a Gaussian distribution ($B=1$, black dashed line) for large system sizes. Disorder values: $W=0.4$ (yellow triangles) and $W=0.8 - 4.0$ in steps of 0.2, where the lines become darker as $W$ increases. (b) Peak positions $L_0$ and peak heights $B_0$, obtained from a scaling collapse of the median and average values of $B$.}
    \label{fig:B_vs_L}
\end{figure}

Fig.~\ref{fig:B_vs_L} shows the $L$-dependence of $B$ for a range of disorder strengths.
In the diffusive phase ($W \lesssim 0.5$), the distribution is Gaussian for all system sizes that we have considered. For intermediate system sizes, in the subdiffusive phase, $B$ initially increases with $L$, indicating that the distributions of matrix elements are becoming less Gaussian. However, when the system size exceeds a critical length $L_0(W)$ this trend reverses, tending back to $B=1$ at large $L$ (non-monotonic $L$-dependences, where a flow initially towards MBL reverses at larger $L$, have been observed before \cite{Serbyn2015criterion} and speak about the nature of the critical point).
The large values of $B$ near $L_0$ originate from broad distributions of $B$, which become narrow again as $L$ becomes large (see the Supplemental Material).

\begin{figure}
    \centering
    \includegraphics[width=\columnwidth]{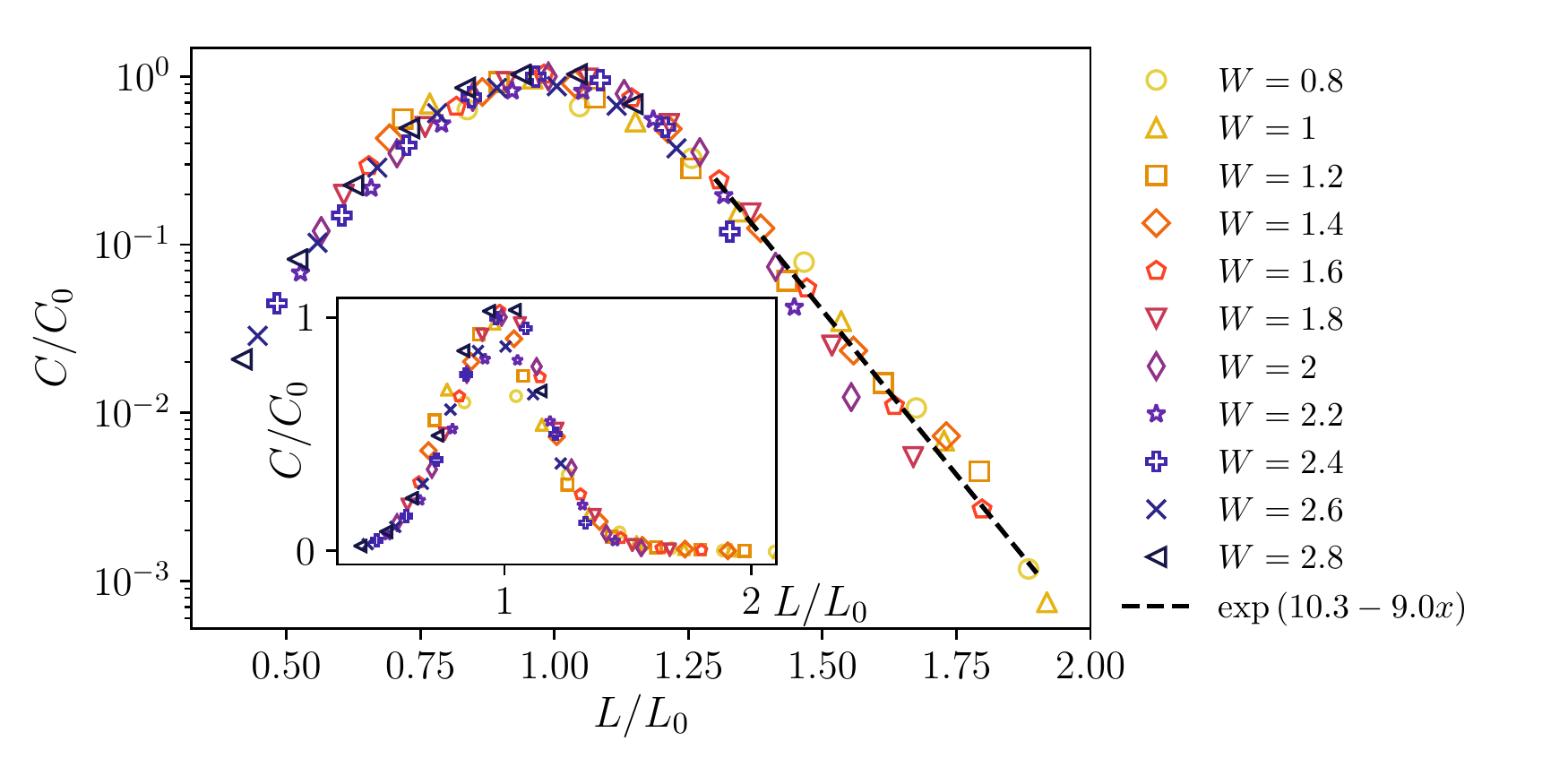}
    \caption{ Numerical collapse of the median value of $C(L,W)$, the inset shows the same data on a normal vertical axis.}
    \label{fig:B_Scaling}
\end{figure}

If the physics of the ergodic region is homogeneous, then one expects that a single function should determine the behaviour of different $L$ and $W$. Therefore we assume that, once $L_0$ and $B_0 := B(L_0(W), W)$ (or equivalently $C_0 = B_0 - 1$) are found, we can write a universal form
\begin{equation}
    C(L,W) = C_0(W) g\left( L / L_0 (W) \right),
\end{equation}
where $g(x)\to 0$ both for $x\to 0,\infty$ and $g(1)=1$.
We perform the collapse numerically, and the extracted values of $L_0(W)$ and $B_0(W)$ are shown in Fig.~\ref{fig:B_vs_L}(b).
The collapse itself is shown in Fig.~\ref{fig:B_Scaling}, and we see that it only becomes imperfect for the largest disorder strengths near the peak, which are the values that we believe may be underestimated due to poor statistics (very large values of $B$ result from heavy-tailed distributions of $R_{\alpha, \beta}$, which require many data points to sample effectively).

$L_0$ is the lengthscale at which resonances that thermalise the system start appearing \cite{Peng2019Comparing}, and it must therefore diverge as $W \to W_c$. One could also define a second length, $L_1(W)$ at which $C\simeq 0$ (say $C<10^{-2}$) or $B\simeq 1$; this would be the length at which resonances become effective ($L_1\gg L_0$).
If one assumes a single parameter scaling, as $W\to W_c^{-}$, then $L_0$ diverges as $L_0(W)=c|W-W_c|^{-\nu}(1+a_1(W-W_c)+...)$. Using the SDRG \cite{dumitrescu2019kosterlitz} (for a quick derivation see the Supplemental Material) one instead finds $L_0(W)\sim {\rm e}^{c|W-W_c|^{-1/2}}$.
We observe the value of $L_0(W)$ to grow linearly with $W$ until the maximum accessible $W=2.8$ (see Fig.~\ref{fig:B_vs_L}(b)), excluding any divergence for $W\lesssim 8$ if we assume $\nu>2$. If we impose $\nu=1$ the critical value is around $W_c\simeq 5$, while for the SDRG prediction the value of $W_c$ compatible with our data is larger, $W_c \approx 6$. Analogously, the value of the maximum $C_0(W)$, which must diverge at the transition $W\to W_c$, shows exponential growth $\ln C_0(W)\simeq W / W_0$ (see Fig.~\ref{fig:B_vs_L}(b)).
The value of $W_0 \simeq 0.4$ is approximately the disorder strength at the diffusion-subdiffusion transition \cite{Znidaric2016Diffusive}.

The exponential growth of $B$ for $L < L_0$ can be understood in the
strong-disorder limit $W \gg 1$ (or equivalently the limit of $L_0 \gg L$).
Here we can apply the usual arguments of perturbative MBL (\emph{e.g.}\ the forward-scattering approximation \cite{ros2015integrals,PhysRevB.93.054201}) to say that the matrix element $R_{\alpha,\beta}$ is independent of $\omega$\footnote{The $\omega$-dependence is not at the exponential level for $\omega=O(1)$.}, $R_{\alpha,\beta}\sim {\rm e}^{-\ell_{\alpha,\beta}/\xi}$,
where $\xi \sim \ln (W / J )^{-1}$ is the LIOMs range or localisation length, and $\ell$ is the number of spin flips connecting $| \beta \rangle$ to $| \alpha \rangle$ to lowest order in perturbation theory.
For typical $\alpha,\beta$, and large $L$, the distance can be taken as an approximately Gaussian-distributed random variable with mean $L/2$ and variance $L/4$.
Computing $\langle R^4\rangle$ and $\langle R^2\rangle$, we find that $B = {\rm e}^{L / \xi^2} / 3$.
The length $\xi$ decreases uniformly with $W$ (see Fig.~\ref{fig:B_vs_L}(a)), therefore representing the effect of local Physics, distinct from that of resonances, which is captured by $L_0$.
The exponential growth with $L$ continues up to $L = L_0 \sim W$, resulting in a maximum value $B_0\sim {\rm e}^{cW}$, explaining the behaviours seen in Fig.~\ref{fig:B_vs_L}(b).

The function $g(x)$ seems to decay exponentially for $x\gtrsim 1$, like $\ln g(x) \sim 10.3-9.0 x$ (the dashed line in Fig.~\ref{fig:B_Scaling}).
This predicts that $L_0 \approx 25$ at the estimated critical disorder strength $W=4$, which is at the upper limit of the system sizes solvable by ED, and that a system of size $L \approx 46$ is required to observe Gaussian-distributed matrix elements.

There are two main messages from this analysis: first, that a scaling form for $B$ works for a wide range of $W,$ and second, that we do not see any sign of the transition in terms of divergence of either $L_0(W)$ or $B_0(W)$ for $W \lesssim 2.8$.
This may suggest that the divergence is found at larger $W$, and our scaling function $g$ works qualitatively fine, but $L_0$ and $B_0$ do not \emph{yet} show signs of the upward curvature necessary for the transition. In this case our $L_0$ and $B_0$, extrapolated as analytic functions of $W$, must be taken as \emph{lower bounds}, and we would require even larger systems than we predict to study the critical point. For example, if we suspect that $W_c\simeq 10$ then we would need at least a system size $L_0 \approx 57$ to study the critical point. 

An alternative possibility is that there is no transition, and that we see only a finite-size effect: if we take a finite system larger than $L_0$ we will always recover the ETH. 
In this sense $L_0(W)$ and $L_1(W)$ would be analogous to the scales observed in the development of ergodicity in $SU(2)$-symmetric random Hamiltonians \cite{protopopov2019non}.
However, in our opinion the data up to $W=2.8$ represent too small an interval of disorder to make such a strong statement; larger $W$ and therefore considerably larger $L$ data are necessary to support this possibility. Additionally, as we will see in the next section, the microscopic timescale for transport shows a qualitative change of its dependence on $W$ at around the same value $W\simeq 2.5$.

\section{Timescales near the MBL transition}

One can argue on purely dynamical grounds that one must be careful when studying the effects of strong disorder on small systems, without resorting to eigenstates and eigenvalues (which in a many-body system due to exponentially small level spacing are anyway not necessarily physically observable).
Namely, for strong disorder the dynamics can become so slow that non-thermodynamic effects (\emph{e.g.}\ finite-size effects) become important before the system has a chance to develop full many-body macroscopic dynamics. More precisely, our analysis relies on two simple facts: (i) to probe the slow asymptotic behaviour of an ergodic phase, one must examine the dynamics at least up to a correspondingly large ``ergodic time'' $t_E$, and (ii) for finite quantum systems it is meaningless to look at times larger than the Heisenberg time $t_H = 2 \pi / \Delta E$. After $t_H$ the dynamics become ``quasiperiodic" (a discrete spectrum is resolved), and are influenced by various non-bulk particularities such as boundary conditions.
Therefore, one can make reliable statements about behaviour in the TDL only if $t_H \gg t_E$! While this condition will always be satisfied in the TDL, in a finite system, and especially for slow dynamics preceding a possible MBL phase, it is not guaranteed.

An important point to keep in mind when generalising quantum chaos concepts, predominantly explored in a single-particle context~\cite{haake1991quantum}, to many-body systems, is that for single-particle (non-localised) systems $t_E$ and $t_H$ are both polynomial in $L$, while in a many-body system $t_H$ is exponentially large in $L$ (as $\Delta E$ is exponentially small). It is largely unexplored how this affects the standard notions of quantum chaos. For example, using level repulsion as an indicator of many-body quantum chaos can lead to a false identification of a system as chaotic~\cite{brenes2018high}, at least if chaoticity is meant to denote a system with sufficiently complex dynamics that, for instance, lead to diffusion.

We shall estimate the ergodic timescale $t_E$, and thereby probe transport, by examining the spreading of an inhomogeneous initial state under unitary time evolution. We note that we can reliably simulate only modest times $t \approx 100$ for rather small disorder strengths $W$. Therefore, we cannot and do not make any claims to find the true asymptotic transport exponents. Nevertheless, we shall use our short-time dynamics to get a gross estimate and show that $t_E$ increases very quickly with disorder strength $W$, which will suffice to make our argument.

\begin{figure}
    \centering
    \includegraphics[width=0.51\columnwidth]{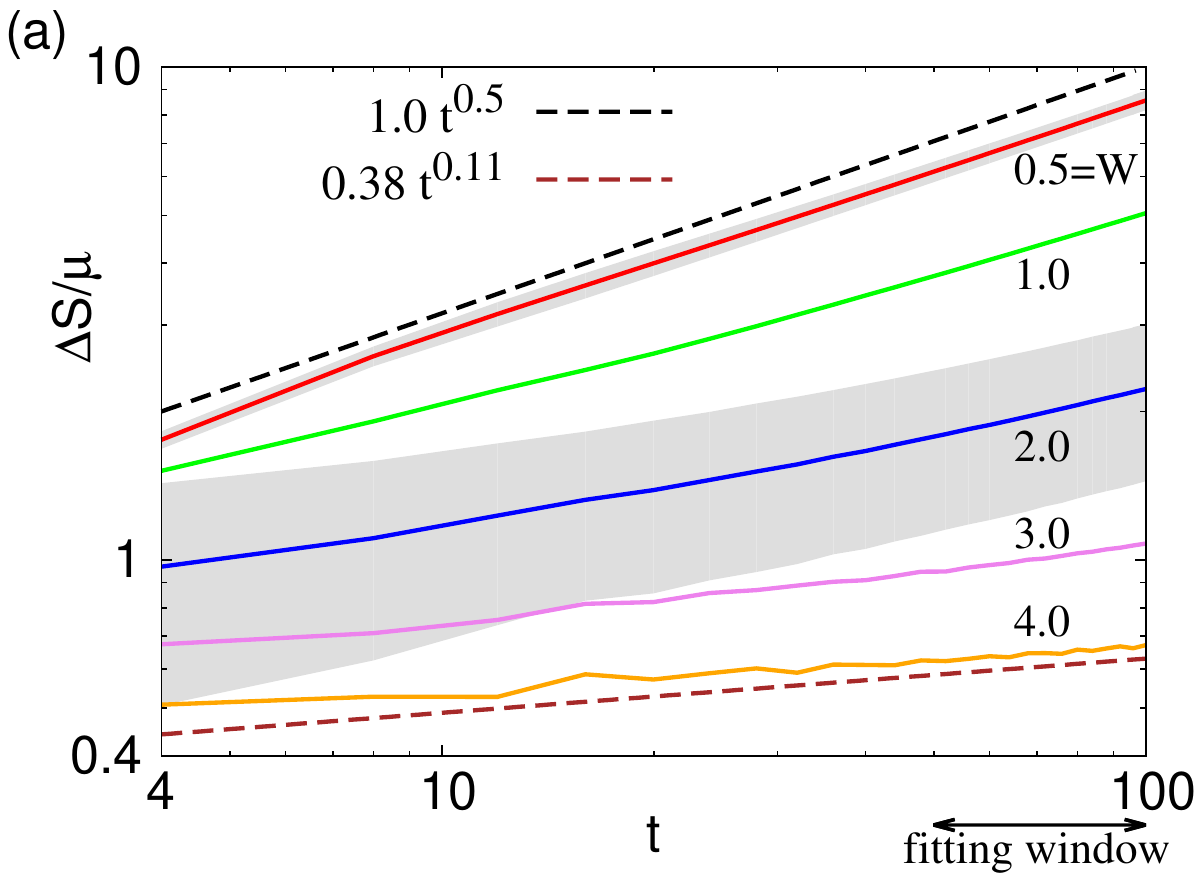}\includegraphics[width=0.48\columnwidth]{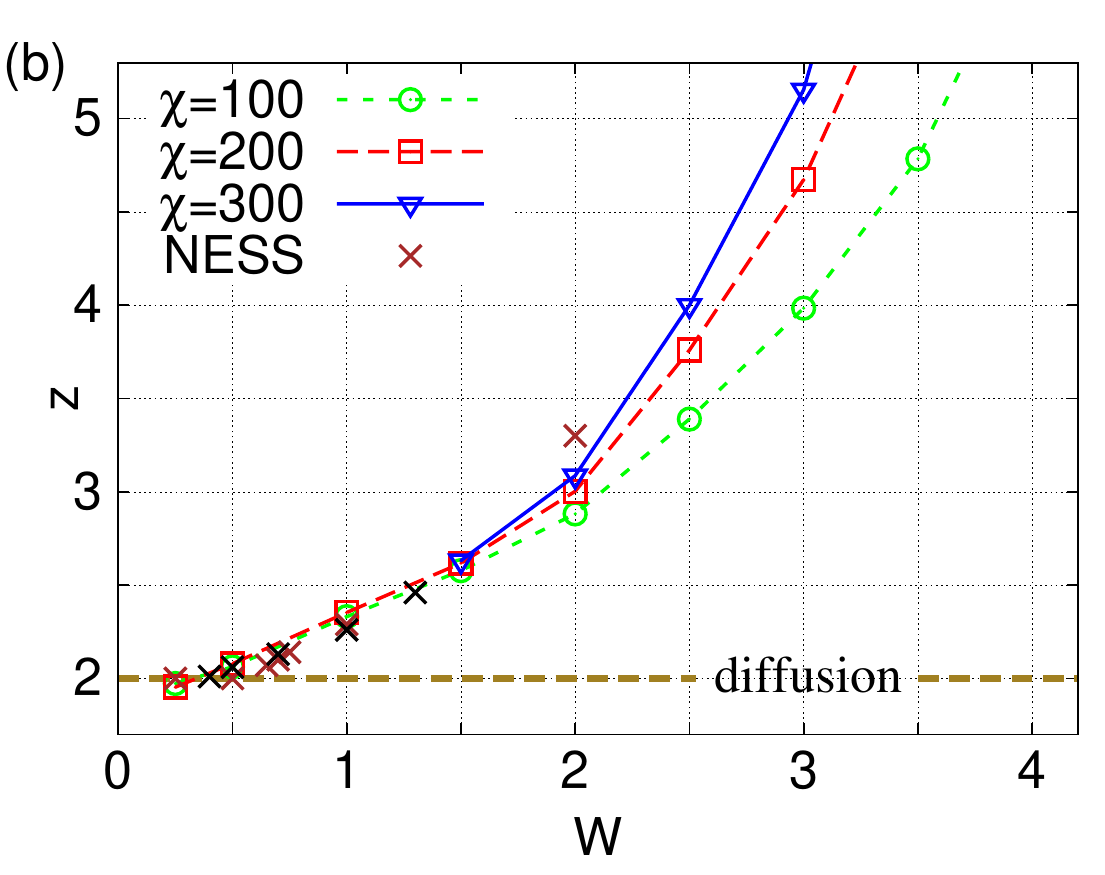}
    \caption{
    Magnetisation transferred across a domain wall under unitary dynamics. (a) Average magnetisation $\Delta S$ as a function of time. The width of the distribution of $\Delta S$ over disorder is indicated by shading for $W=0.5, 2$.
    (b) Finite-time scaling exponent $z$ from the data shown in the left panel (fitting window $t\in[50,100]$), obtained for several different MPO bond dimensions $\chi$ and from NESS exponents in boundary-driven systems~\cite{Znidaric2016Diffusive,Schulz2019Phenomenology}.}
    \label{fig:z}
\end{figure}

As an initial state we take a weakly polarised domain wall, $\rho(0) \sim \prod_{k=1}^{L/2} (1+\mu\sigma_k^z) \otimes \prod_{k=\frac{L}{2}+1}^L \!(1-\mu \sigma_k^z)$, with a small polarisation $\mu=10^{-3}$. Such an initial state is useful for probing infinite-temperature spin transport~\cite{pal2010MBL,ljubotina2017spin}. The initial state is evolved in time as $\rho(t)={\rm e}^{-i H t}\rho(0) {\rm e}^{i H t}$, with $\rho$ encoded using a matrix product operator (MPO) ansatz of bond dimension $\chi$, and using tDMRG with a 4\textsuperscript{th} order Trotter-Suzuki timestep $\Delta t=0.2$. To quantify the slowness of the domain wall decay we calculate the transferred magnetisation across the mid-point up to a time $t$, $\Delta S(t)=\mu \frac{L}{2}-\sum_{k=1}^{L/2} {\rm tr}{[\rho(t)\sigma_k^z]}$. The dimensionless quantity $\Delta S(t)/\mu$ then tells us how many spins have moved after a time $t$. In Fig.~\ref{fig:z}(a) we show the evolution of the disorder-averaged $\Delta S(t)$ for a few representative disorder strengths.
It is clear that the dynamics get slower with $W$.
The method's limitation is the maximal time one can reach with a given MPO bond dimension $\chi$, while on the other hand we always take a sufficiently large $L$ such that there are no boundary effects (in practice this means sizes $L=32-120$, averaging over $10-560$ disorder samples). Note how the widths of the distributions of $\Delta S$, shown by shading for $W=0.5$ and $W=2$, increase with $W$.

From Fig.~\ref{fig:z}(a) it appears that the magnetisation relaxes in a power-law fashion,
\begin{equation}
  \Delta S(t)/\mu \approx A t^{1/z}.
  \label{eq:fit}
\end{equation}
The fitted $A$ and $z$ are shown in Fig.~\ref{fig:z}(b). It is clear that due to limited times, at least for larger $W$, one cannot claim that the so-obtained $z$ is the correct asymptotic transport coefficient (for instance, at $W=3$ only about $1$ spin is transferred between $t=10$ and $t=100$), nevertheless, it is clear that the dynamics at such $W$ are very slow. At small $W$ we find good agreement with the results from boundary-driven open systems on the scaling of the non-equilibrium steady state (NESS) current with system size, $j \sim L^{1-z}$ \cite{Znidaric2016Diffusive,Schulz2019Phenomenology}.

We also note that we cannot determine the point where $z$ diverges (\emph{i.e.}\ the MBL transition). First, while the data has converged with $\chi$ for small $W$, this is not the case for $W>2$ where the numerics are harder. However, we see a systematic tendency for $z$ to increase with $\chi$, so our $z(\chi=300)$ can be used as a lower bound. Second, the exponent $z$ tends to decrease with time. For instance, at $W=2$ and $\chi=300$ we get exponents $1/z=0.264, 0.300, 0.324$ using fitting windows $t\in [20,40],[40,60],[60,80]$ respectively. Using longer times would therefore tend to move a possible divergence of $z$ towards larger $W$.

From the fitted $A(W)$ and $z(W)$ values we can now estimate the time at which a given number of spins will be transferred across the middle of the chain.
This can then serve as an estimate for the ergodic time.
We denote by $t_N$ the time when $\Delta S/\mu=N$, \emph{i.e.}\ the time when $N$ spins are transported across the domain wall, which is asymptotically equal to $t_N=(N/A)^z$ according to \eqref{eq:fit}. Depending on the context different values of $N$ might be used as a criterion for the ergodic time. To assess many-body transport in the TDL one certainly requires $N \gg 1$ to be able to discuss dynamics of many particles, and to presumably reach a limit where some coarse-grained hydrodynamic description applies. In a finite system a reasonable value might be a fixed fraction of spins, \emph{e.g.}\ $N/L=1/4$, to probe relaxation on the largest wavelength.

\begin{figure}
    \centering
       \includegraphics[width=0.51\columnwidth]{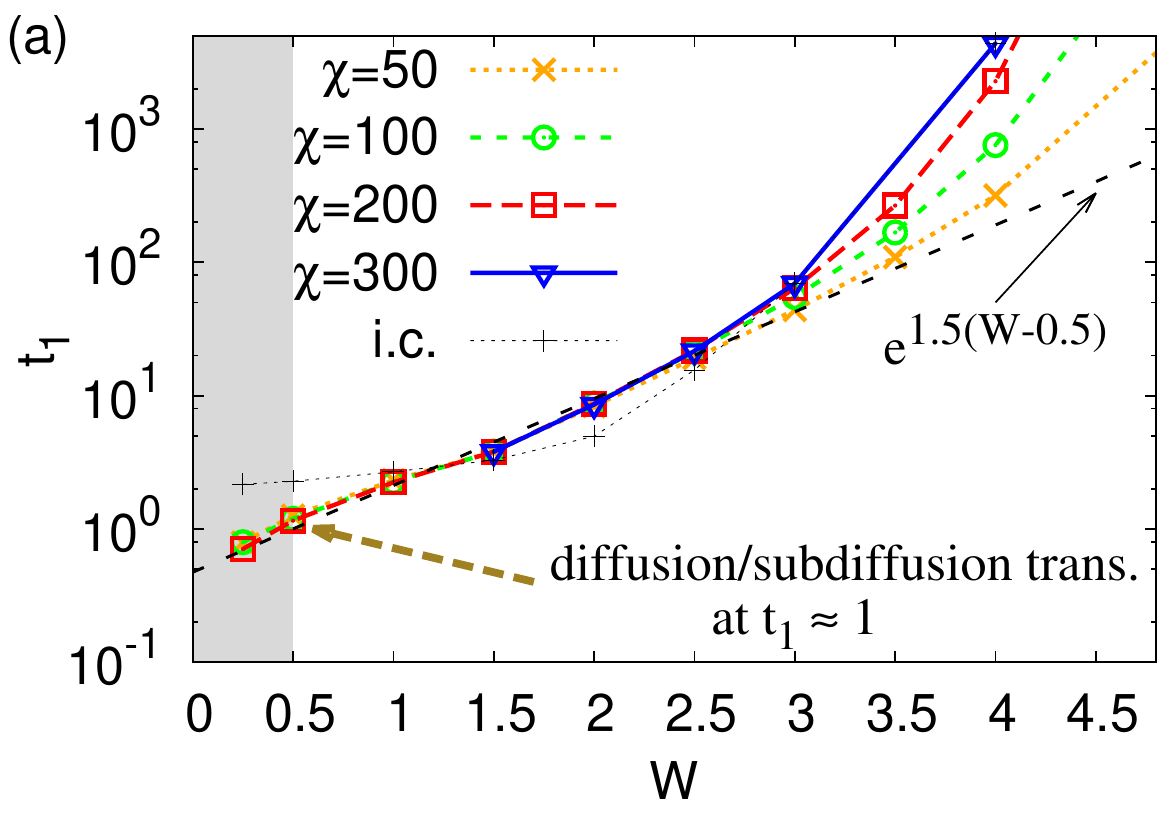}\includegraphics[width=0.48\columnwidth]{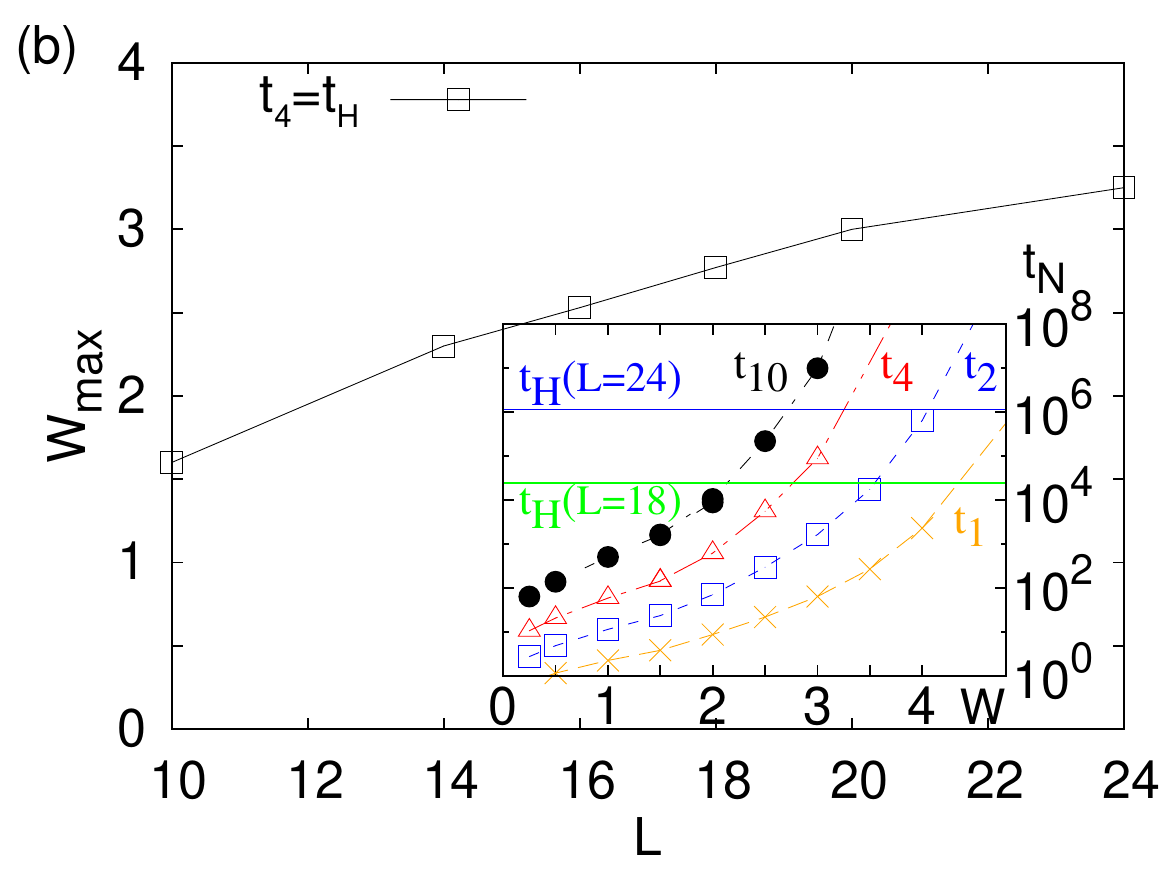}
    \caption{
    Timescales in the disordered Heisenberg model.
    (a) Growth of the asymptotic time $t_1$ needed to move one spin with increasing $W$, shown for several bond dimensions $\chi$. Pluses labelled ``i.c.'' denote $t_1$ determined for our particular initial state, (\emph{i.e.}\ not from $A$ and $z$).
    (b) Main panel: Maximal $W$ for a system of size $L$ below which at least 4 spins are transferred at $t_H$.
    Inset: Time $t_N$ as a function of $W$ using data with $\chi=200-300$. Horizontal lines indicate $t_H$ for $L=18,24$.
    }
    \label{fig:timescales}
\end{figure}

Let us first analyse the time $t_1(W)$ in which a single spin moves. Using it one can write
\begin{equation}
  t_N(W)=N^z \, t_1(W),
  \label{eq:tN}
\end{equation}
that is, $t_N$ is simply some high multiple (for large $z$) of $t_1$. Fig.~\ref{fig:timescales}(a) shows the empirically determined $t_1$ using the fitted values of $A$ and $z$ (\emph{i.e.}\ $t_1=(1/A)^z$); note that because $t_1$ is obtained from $A$ and $z$, for small $W$ it is not equal to the time at which the first spin crosses the mid-point from a particular initial state (pluses in Fig.~\ref{fig:timescales}(a)).

We can see that $t_1$ is a very steep function of $W$, in fact roughly compatible with an exponential dependence $t_1 \sim {\rm e}^{W/\tilde W}$ for a range of intermediate $W$ (a similar exponential slowing-down with $W$ has also been observed in Refs.~\cite{Barisic2010Conductivity,Schiulaz2019Thouless,vsuntajs2019quantum}). This is the crux of the problem! If the MBL transition happens at large $W$, and in all claimed MBL systems this is the case, then in small systems it can easily happen that $t_1$, and even more so $t_N$, is in fact larger than the Heisenberg time $t_H$. Below we compare explicit numbers for the XXX spin chain.

We note that the numerical value of $\tilde W\simeq 0.66$ is similar to the parameter extracted from the Binder cumulant analysis ($W_0 \approx 0.4$), and it is also compatible with the value of the disorder at which the transition from diffusive transport happens. At $W\simeq W_c^{\rm sub} \approx 0.5$ we have $t_1 \approx 1$, suggesting that subdiffusion emerges when the natural timescale of the effective dynamics becomes longer than the bare hopping timescale ($1/J=1$ in our units).

In Fig.~\ref{fig:timescales}(a) one can also see that for $W \gtrsim 2-2.5$, when $t_1 \gtrsim 20 $, the $W$-dependence of $t_1$ becomes faster than exponential. At $W>2.5$ we also observe (data not shown) that there is a visible non-zero probability for an individual disorder realisation to result in an essentially zero value of $1/z$. As a result, the distribution of $z$ gets fat-tailed simply due to the Jacobian, $p(z)=p(x:=1/z)|dx/dz|$ where $|dx/dz|=1/z^2$. In turn, fluctuations in $z$ between realisations become large, and the results depend on which quantity one averages (\emph{e.g.}\ the mean is not the same as the median). While it is tempting to speculate that this is a sign of a possible transition (Griffiths effects, etc.), we again stress that to give such a statement a sound foundation one would have to simulate longer times: to distinguish slow transport from no transport one needs large times and correspondingly large systems -- the very point we are trying to make.

Let us now take the above $t_1$ and some reasonable $N$, and make an estimate for the ergodic time $t_E \sim t_N=N^z t_1$. We estimate the Heisenberg time $t_{\rm H}=2\pi/\Delta E$ from the mean level spacing $\Delta E$ in the centre of a half-filling sector, resulting in $t_{\rm H}\approx \frac{2^L}{L W}\frac{8}{\sqrt{4/3+3/W^2}}$. For $W=3$ this gives a Heisenberg time of $t_H \approx 2\cdot 10^3$ for $L=14$, and $t_H\approx 2\cdot 10^4$ for $L=18$. On the other hand, for $W=3$ we have $t_1 \sim 10^2$ and $z>5$, so $t_3 = 3^z t_1 \gtrsim t_{\rm H}(L=18,W=3)$. Observing the dynamics of even 3 spins is impossible with $L=18$ and $W \gtrsim 3$ because the Heisenberg time ``kicks in" before those three spins have had time to move!

In the inset of Fig.~\ref{fig:timescales}(b) we show a more detailed comparison. We can for instance see that with $L=18$ (the typical maximal size used in ED studies) $t_{\rm H}$ is barely long enough to transfer $N=4$ spins at $W\approx 2.7$. When $W$ is larger, $t_H$ becomes smaller than $t_4$ and statements about the behaviour in the TDL are questionable. For $L=24$ and requiring $N=4$ one can get only up to $W \approx 3.2$ (considering that our $t_N$ are only a lower limit estimate due to finite $\chi$, the true value of $W$ is likely even smaller). This means that for $W>3.2$ even $L=24$ is not large enough to resolve the slow dynamics. On a similar note, in recent experiments \cite{Bloch2015} one has access to times of order $t \sim 100$, so beyond $W\approx 2$ one cannot really distinguish localisation from slow transport (only about $N=2$ spins are transferred at $t=100$ and $W=2$). The main panel of Fig.~\ref{fig:timescales}(b) shows the largest $W$ that can be studied in a system of size $L$, demanding that $t_4  < t_{\rm H}$. Finally, we observe that the exponential dependence of $t_1$ on $W$ for $W \lesssim 2.5$ means that the minimal size grows linearly with $W$ in that range of $W$, compatible with a similar finding for the Binder cumulant.

\section{Conclusions}

More than 10 years after the first works proposing the existence of an MBL phase, and of an MBL-ETH transition, the transition itself has resisted all attacks aimed at defining its critical properties. We have asked why this is the case, and in particular what are the length- and timescales necessary to enter an asymptotic region. We have found that the lengthscales for an exact diagonalisation study must grow at least linearly in $W$, and that if the critical disorder is around $W=10$ we need system sizes of at least 50 spins to check the hypothesis against the null hypothesis, \emph{i.e.}\ that no transition exists and everything is just finite-size corrections to ETH Physics. In particular, the value of the Binder cumulant at the peak of the finite size curve scales exponentially with the disorder $W$, with a scale that is equal, within errors, to the disorder strength at which one finds the diffusion-subdiffusion transition.
We have also looked at the time taken for one or more spins to diffuse away from a domain wall configuration. This time also increases exponentially with $W$, with approximately the same rate as the Binder cumulant. For finite systems, as the disorder strength increases this time becomes rapidly larger than the Heisenberg time when the discrete spectrum is recognised.

More work is needed to distinguish between the possibilities that $W_c > 4$ or that the MBL transition does not occur in this particular model Hamiltonian. However, we think we have made a clear point that this cannot be done with simply further numerical works or experiments at marginally larger system sizes or times than what the current technology can provide.

\acknowledgements
The authors thank D.~Abanin, M.~Dalmonte, and S.~Parameswaran for useful conversations. This work was supported by the Trieste Institute for the Theory of Quantum Technologies (RKP, AS, MS, SRT), CINECA ISCRA grant: project IsC66 EDNAS, and Grant No.~J1-1698 and program No. P1-0402 of the Slovenian Research Agency.

\bibliography{MBLbib}

\begin{thebibliography}{10}
\expandafter\ifx\csname url\endcsname\relax\def\url#1{\texttt{#1}}\fi

\bibitem{landau2013course}
\Name{Landau L.~D. \and Lifshitz E.~M.} \Book{Course of theoretical physics}
  (Elsevier) 2013.

\bibitem{rigol2008thermalization}
\Name{Rigol M., Dunjko V. \and Olshanii M.} \REVIEW{Nature}{452}{2008}{854}.

\bibitem{Goldstein2015}
\Name{Goldstein S., Huse D.~A., Lebowitz J.~L. \and Tumulka R.} \REVIEW{Phys.
  Rev. Lett.}{115}{2015}{100402}.

\bibitem{Basko:2006hh}
\Name{Basko D.~M., Aleiner I.~L. \and Altshuler B.~L.} \REVIEW{Ann.
  Phys.}{321}{2006}{1126}.

\bibitem{gornyi2005interacting}
\Name{Gornyi I., Mirlin A. \and Polyakov D.} \REVIEW{Phys. Rev.
  Lett.}{95}{2005}{206603}.

\bibitem{de2013ergodicity}
\Name{{De Luca} A. \and Scardicchio A.} \REVIEW{Europhys.
  Lett.}{101}{2013}{37003}.

\bibitem{huse2014phenomenology}
\Name{Huse D.~A., Nandkishore R. \and Oganesyan V.} \REVIEW{Phys. Rev.
  B}{90}{2014}{174202}.

\bibitem{nandkishore2015many}
\Name{Nandkishore R. \and Huse D.~A.} \REVIEW{Annu. Rev. Condens. Matter
  Phys.}{6}{2015}{15}.

\bibitem{luitz2015many}
\Name{Luitz D.~J., Laflorencie N. \and Alet F.} \REVIEW{Phys. Rev.
  B}{91}{2015}{081103}.

\bibitem{abanin2017recent}
\Name{Abanin D.~A. \and Papi{\'c} Z.} \REVIEW{Annalen der
  Physik}{529}{2017}{1700169}.

\bibitem{zhang2017observation}
\Name{Zhang J., Hess P., Kyprianidis A., Becker P., Lee A., Smith J., Pagano
  G., Potirniche I.-D., Potter A.~C., Vishwanath A. \etal}
  \REVIEW{Nature}{543}{2017}{217}.

\bibitem{sacha2017time}
\Name{Sacha K. \and Zakrzewski J.} \REVIEW{Rep. Prog. Phys.}{81}{2017}{016401}.

\bibitem{Abanin2016Theory}
\Name{Abanin D.~A., Roeck W.~D. \and Huveneers F.} \REVIEW{Annals of
  Physics}{372}{2016}{1}.

\bibitem{brenes2018many}
\Name{Brenes M., Dalmonte M., Heyl M. \and Scardicchio A.} \REVIEW{Phys. Rev.
  Lett.}{120}{2018}{030601}.

\bibitem{Papic2015}
\Name{Papi{\'c} Z., Stoudenmire E.~M. \and Abanin D.~A.} \REVIEW{Ann.
  Phys}{362}{2015}{714}.

\bibitem{schiulaz2014glass}
\Name{Schiulaz M. \and M\"uller M.} \REVIEW{AIP Conf. Proc.}{1610}{2014}{11}.

\bibitem{pino2015metallic}
\Name{Pino M., Ioffe L.~B. \and Altshuler B.~L.} \REVIEW{Proc. Natl. Acad. Sci.
  U.S.A.}{113}{2015}{536}.

\bibitem{yao2016quasi}
\Name{Yao N.~Y., Laumann C.~R., Cirac J.~I., Lukin M.~D. \and Moore J.~E.}
  \REVIEW{Phys. Rev. Lett.}{117}{2016}{240601}.

\bibitem{nandkishore2017amany}
\Name{Nandkishore R.~M. \and Sondhi S.~L.} \REVIEW{Phys. Rev.
  X}{7}{2017}{041021}.

\bibitem{Schulz2019Stark}
\Name{Schulz M., Hooley C.~A., Moessner R. \and Pollmann F.} \REVIEW{Phys. Rev.
  Lett.}{122}{2019}{040606}.

\bibitem{vanNieuwenburg2019From}
\Name{van Nieuwenburg E., Baum Y. \and Refael G.} \REVIEW{Proc. Natl. Acad.
  Sci. U.S.A.}{116}{2019}{9269}.

\bibitem{serbyn2013local}
\Name{Serbyn M., Papi{\'c} Z. \and Abanin D.~A.} \REVIEW{Phys. Rev
  Lett.}{111}{2013}{127201}.

\bibitem{imbrie2014many}
\Name{{Imbrie} J.~Z.} \REVIEW{J. Stat. Phys.}{163}{2016}{998}.

\bibitem{ros2015integrals}
\Name{Ros V., M{\"u}ller M. \and Scardicchio A.} \REVIEW{Nucl. Phys.
  B}{891}{2015}{420}.

\bibitem{imbrie2017local}
\Name{Imbrie J.~Z., Ros V. \and Scardicchio A.} \REVIEW{Annalen der
  Physik}{529}{2017}{1600278}.

\bibitem{znidaric08}
\Name{{\v Znidari\v c} M., Prosen T. \and {Prelov\v sek} P.}
  \REVIEW{Phys.~Rev.~B}{77}{2008}{064426}.

\bibitem{Reichman2014Absence}
\Name{{Bar Lev} Y., Cohen G. \and Reichman D.~R.} \REVIEW{Phys. Rev.
  Lett.}{114}{2015}{100601}.

\bibitem{Varma2017Energy}
\Name{Varma V.~K., Lerose A., Pietracaprina F., Goold J. \and Scardicchio A.}
  \REVIEW{J. Stat. Mech.: Theory Exp.}{2017}{2017}{053101}.

\bibitem{AgarwalAnomalousDiffusion}
\Name{Agarwal K., Gopalakrishnan S., Knap M., M{\"u}ller M. \and Demler E.}
  \REVIEW{Phys. Rev. Lett.}{114}{2015}{160401}.

\bibitem{Znidaric2016Diffusive}
\Name{\v{Z}nidari\v{c} M., Scardicchio A. \and Varma V.~K.} \REVIEW{Phys. Rev.
  Lett.}{117}{2016}{040601}.

\bibitem{Schulz2018Energy}
\Name{Schulz M., Taylor S.~R., Hooley C.~A. \and Scardicchio A.} \REVIEW{Phys.
  Rev. B}{98}{2018}{180201}.

\bibitem{Mendoza2019Asymmetry}
\Name{Mendoza-Arenas J.~J., \ifmmode \check{Z}\else
  \v{Z}\fi{}nidari\ifmmode~\check{c}\else \v{c}\fi{} M., Varma V.~K., Goold J.,
  Clark S.~R. \and Scardicchio A.} \REVIEW{Phys. Rev. B}{99}{2019}{094435}.

\bibitem{Schulz2019Phenomenology}
\Name{Schulz M., Taylor S.~R., Scardicchio A. \and \v{Z}nidari\v{c} M.}
  \REVIEW{arXiv preprint arXiv:1909.09507}{}{2019}{}.

\bibitem{Bloch2015}
\Name{{Schreiber} M., {Hodgman} S.~S., {Bordia} P., {L{\"u}schen} H.~P.,
  {Fischer} M.~H., {Vosk} R., {Altman} E., {Schneider} U. \and {Bloch} I.}
  \REVIEW{Science}{349}{2015}{842}.

\bibitem{Khemani2017Critical}
\Name{Khemani V., Lim S.~P., Sheng D.~N. \and Huse D.~A.} \REVIEW{Phys. Rev.
  X}{7}{2017}{021013}.

\bibitem{vsuntajs2019quantum}
\Name{{\v{S}}untajs J., Bon{\v{c}}a J., Prosen T. \and Vidmar L.} \REVIEW{arXiv
  preprint arXiv:1905.06345}{}{2019}{}.

\bibitem{Abanin2019Distinguishing}
\Name{Abanin D.~A., Bardarson J.~H., De~Tomasi G., Gopalakrishnan S., Khemani
  V., Parameswaran S.~A., Pollmann F., Potter A.~C., Serbyn M. \and Vasseur R.}
  \REVIEW{arXiv preprint arXiv:1911.04501}{}{2019}{}.

\bibitem{Sierant2019Thouless}
\Name{Sierant P., Delande D. \and Zakrzewski J.} \REVIEW{arXiv preprint
  arXiv:1911.06221}{}{2019}{}.

\bibitem{pietracaprina2018shift}
\Name{Pietracaprina F., Mac{\'e} N., Luitz D.~J. \and Alet F.} \REVIEW{SciPost
  Phys.}{5}{2018}{45}.

\bibitem{alet2015}
\Name{Luitz D.~J., Laflorencie N. \and Alet F.} \REVIEW{Phys. Rev.
  B}{91}{2015}{081103}.

\bibitem{Pietracaprina2017Entanglement}
\Name{Pietracaprina F., Parisi G., Mariano A., Pascazio S. \and Scardicchio A.}
  \REVIEW{J. Stat. Mech.: Theory Exp.}{2017}{2017}{113102}.

\bibitem{Mace2019Multifractal}
\Name{Mac\'e N., Alet F. \and Laflorencie N.} \REVIEW{Phys. Rev.
  Lett.}{123}{2019}{180601}.

\bibitem{chandran2015finite}
\Name{Chandran A., Laumann C. \and Oganesyan V.} \REVIEW{arXiv:
  1509.04285}{}{2015}{}.

\bibitem{Altman2015}
\Name{Altman E. \and Vosk R.} \REVIEW{Annu. Rev. Condens. Matter
  Phys.}{6}{2015}{383}.

\bibitem{AltmanTheory2015}
\Name{Vosk R., Huse D.~A. \and Altman E.} \REVIEW{Phys. Rev.
  X}{5}{2015}{031032}.

\bibitem{dumitrescu2017scaling}
\Name{Dumitrescu P.~T., Vasseur R. \and Potter A.~C.} \REVIEW{Phys. Rev.
  Lett.}{119}{2017}{110604}.

\bibitem{thiery2018many}
\Name{Thiery T., Huveneers F., M{\"u}ller M. \and De~Roeck W.} \REVIEW{Phys.
  Rev. Lett.}{121}{2018}{140601}.

\bibitem{dumitrescu2019kosterlitz}
\Name{Dumitrescu P.~T., Goremykina A., Parameswaran S.~A., Serbyn M. \and
  Vasseur R.} \REVIEW{Phys. Rev. B}{99}{2019}{094205}.

\bibitem{goremykina2019analytically}
\Name{Goremykina A., Vasseur R. \and Serbyn M.} \REVIEW{Phys. Rev.
  Lett.}{122}{2019}{040601}.

\bibitem{Luitz2016Anomalous}
\Name{Luitz D.~J. \and Bar~Lev Y.} \REVIEW{Phys. Rev.
  Lett.}{117}{2016}{170404}.

\bibitem{colmenarez2019statistics}
\Name{Colmenarez L., McClarty P.~A., Haque M. \and Luitz D.~J.} \REVIEW{SciPost
  Phys.}{7}{2019}{64}.

\bibitem{Foini2019Eigenstate}
\Name{Foini L. \and Kurchan J.} \REVIEW{Phys. Rev. Lett.}{123}{2019}{260601}.

\bibitem{Luitz2019Multifractality}
\Name{Luitz D.~J., Khaymovich I. \and Lev Y.~B.} \REVIEW{arXiv preprint
  arXiv:1909.06380}{}{2019}{}.

\bibitem{Deutsch1991}
\Name{Deutsch J.~M.} \REVIEW{Phys. Rev. A}{43}{1991}{2046}.

\bibitem{Srednicki1994}
\Name{Srednicki M.} \REVIEW{Phys. Rev. E}{50}{1994}{888}.

\bibitem{Berry1977Regular}
\Name{Berry M.~V.} \REVIEW{J. Phys. A}{10}{1977}{2083}.

\bibitem{EDData}
Contact the corresponding author to access the raw data.

\bibitem{abhyankar2018petsc}
\Name{Abhyankar S., Brown J., Constantinescu E.~M., Ghosh D., Smith B.~F. \and
  Zhang H.} \REVIEW{arXiv preprint arXiv:1806.01437}{}{2018}{}.

\bibitem{petsc-efficient}
\Name{Balay S., Gropp W.~D., McInnes L.~C. \and Smith B.~F.} \Book{Efficient
  management of parallelism in object oriented numerical software libraries} in
  proc. of \Book{Modern Software Tools in Scientific Computing}, edited by
  \Name{Arge E., Bruaset A.~M. \and Langtangen H.~P.} (Birkh{\"{a}}user Press)
  1997 pp. 163--202.

\bibitem{SLEPc}
\Name{Hernandez V., Roman J.~E. \and Vidal V.} \REVIEW{ACM Trans. Math.
  Softw.}{31}{2005}{351}.

\bibitem{Hernandez2003SLEPc}
\Name{Hernandez V., Roman J.~E. \and Vidal V.} \REVIEW{Lect. Notes Comput.
  Sci.}{2565}{2003}{377}.

\bibitem{slepc-users-manual}
\Name{Roman J.~E., Campos C., Romero E. \and Tomas A.} Tech. Rep. D. Sistemes
  Inform\`atics i Computaci\'o, Universitat Polit\`ecnica de Val\`encia
  DSIC-II/24/02 - Revision 3.11 (2019).

\bibitem{MUMPs_1}
\Name{Amestoy P.~R., Duff I.~S., Koster J. \and L'Excellent J.-Y.} \REVIEW{SIAM
  Journal on Matrix Analysis and Applications}{23}{2001}{15}.

\bibitem{MUMPs_2}
\Name{Amestoy P.~R., Guermouche A., L'Excellent J.-Y. \and Pralet S.}
  \REVIEW{Parallel Computing}{32}{2006}{136}.

\bibitem{Serbyn2015criterion}
\Name{Serbyn M., Papi{\'c} Z. \and Abanin D.~A.} \REVIEW{Phys. Rev.
  X}{5}{2015}{041047}.

\bibitem{Peng2019Comparing}
\Name{Peng P., Li Z., Yan H., Wei K.~X. \and Cappellaro P.} \REVIEW{Phys. Rev.
  B}{100}{2019}{214203}.

\bibitem{PhysRevB.93.054201}
\Name{Pietracaprina F., Ros V. \and Scardicchio A.} \REVIEW{Phys. Rev.
  B}{93}{2016}{054201}.

\bibitem{protopopov2019non}
\Name{Protopopov I.~V., Panda R.~K., Parolini T., Scardicchio A., Demler E.
  \and Abanin D.~A.} \REVIEW{arXiv preprint arXiv:1902.09236}{}{2019}{}.

\bibitem{haake1991quantum}
\Name{Haake F.} \Book{Quantum signatures of chaos} in \Book{Quantum Coherence
  in Mesoscopic Systems} (Springer) 1991 pp. 583--595.

\bibitem{brenes2018high}
\Name{Brenes M., Mascarenhas E., Rigol M. \and Goold J.} \REVIEW{Phys. Rev.
  B}{98}{2018}{235128}.

\bibitem{pal2010MBL}
\Name{Pal A. \and Huse D.~A.} \REVIEW{Phys. Rev. B}{82}{2010}{174411}.

\bibitem{ljubotina2017spin}
\Name{Ljubotina M., {\v{Z}}nidari{\v{c}} M. \and Prosen T.} \REVIEW{Nat.
  Commun.}{8}{2017}{16117}.

\bibitem{Barisic2010Conductivity}
\Name{Bari\ifmmode \check{s}\else \v{s}\fi{}i\ifmmode~\acute{c}\else \'{c}\fi{}
  O.~S. \and Prelov\ifmmode~\check{s}\else \v{s}\fi{}ek P.} \REVIEW{Phys. Rev.
  B}{82}{2010}{161106}.

\bibitem{Schiulaz2019Thouless}
\Name{Schiulaz M., Torres-Herrera E.~J. \and Santos L.~F.} \REVIEW{Phys. Rev.
  B}{99}{2019}{174313}.

\end{thebibliography}
\bibliographystyle{eplbib}

\clearpage
\begin{center}
\textbf{\Large{Supplemental Material}}
\end{center}

\section{Additional Binder cumulant data}
\begin{figure}[h!]
    \centering
    \includegraphics[width=\columnwidth]{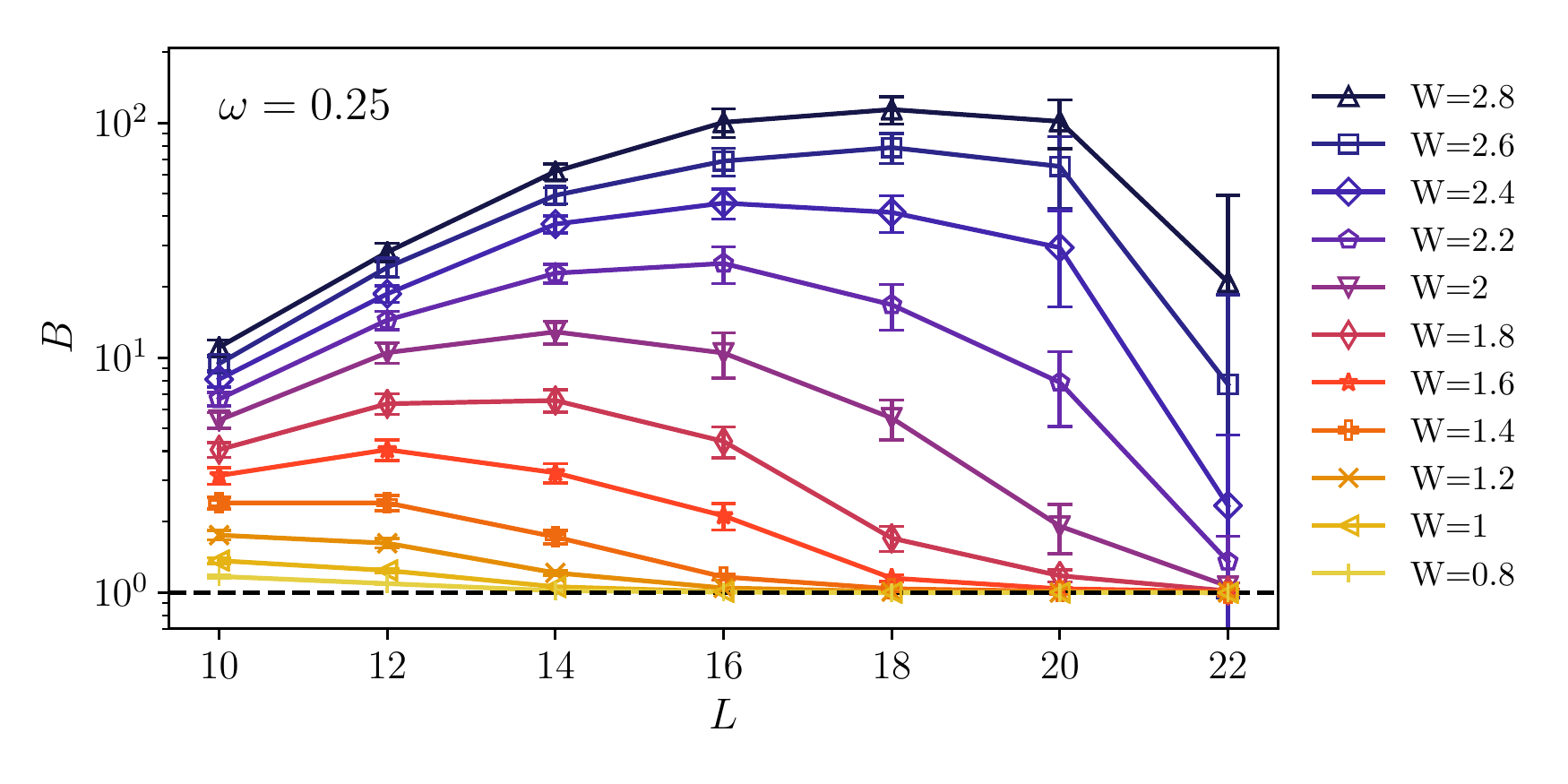}
    \includegraphics[width=\columnwidth]{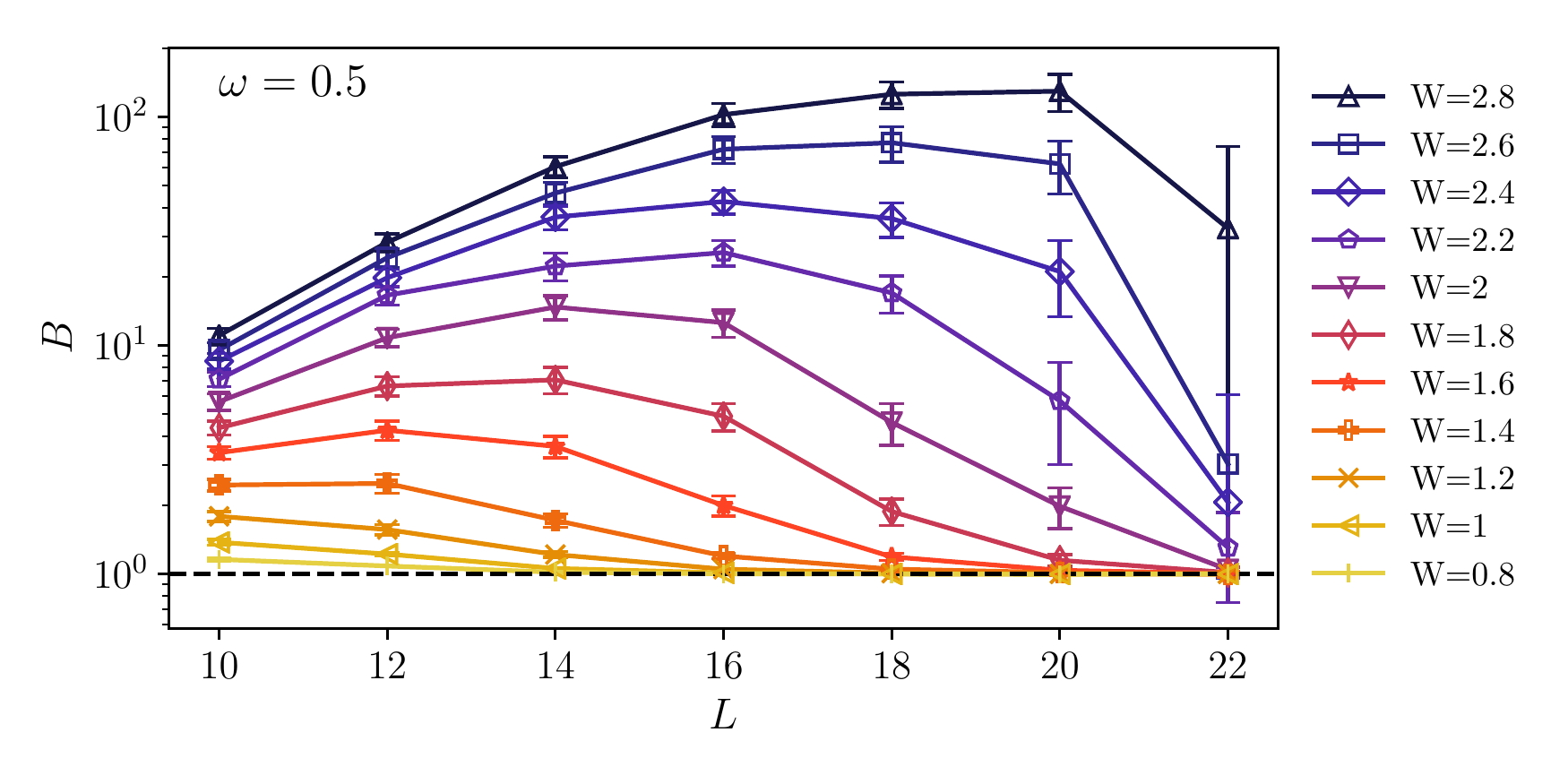}
    \includegraphics[width=\columnwidth]{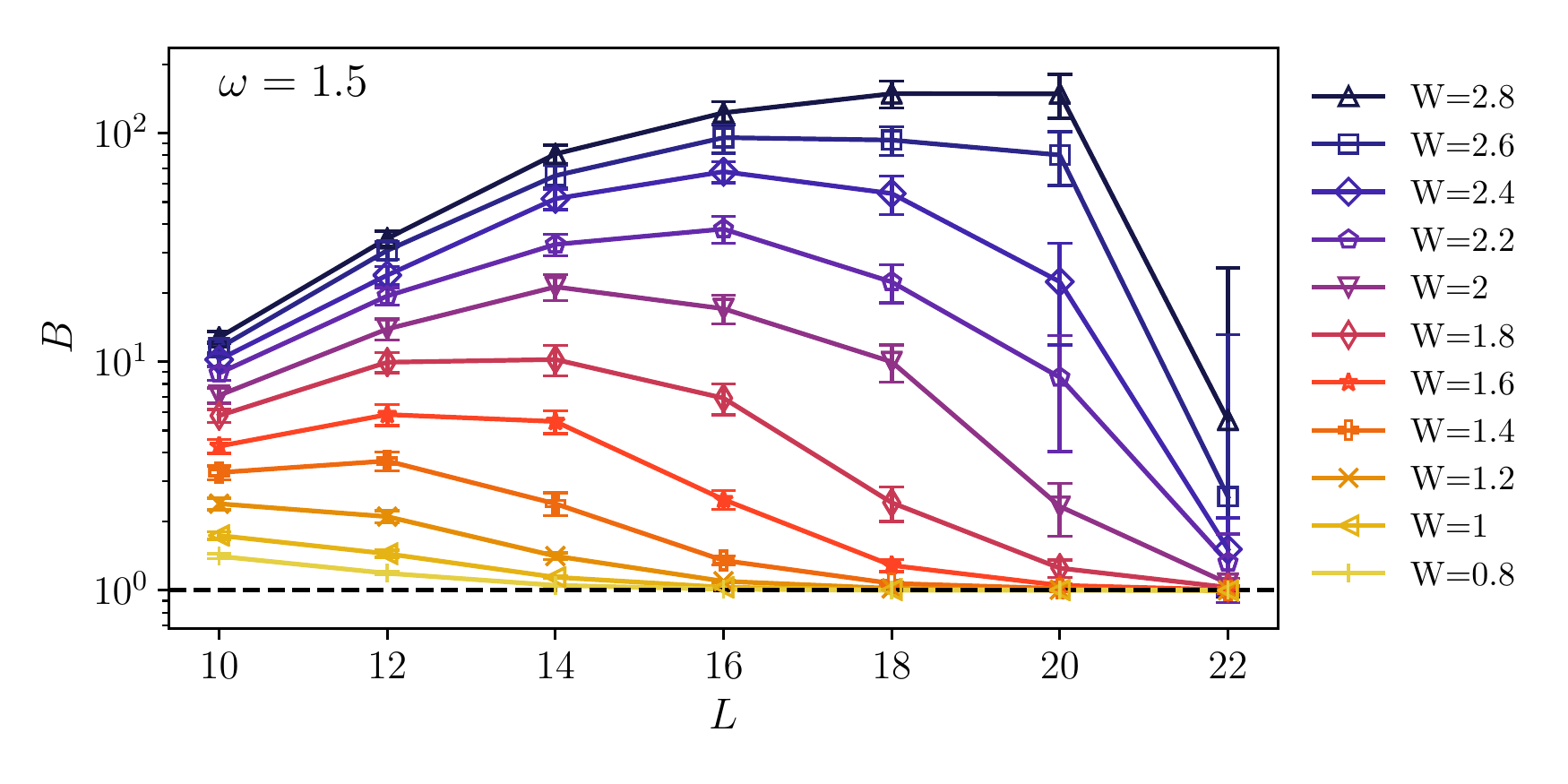}
    \caption{The median Binder cumulant $B$ as a function of system size $L$ for several disorder strengths $W$, shown for $\omega = 0.25$ (top), $\omega = 0.5$ (middle), and $\omega = 1.5$ (bottom).}
    \label{fig:MoreBs}
\end{figure}

We find no qualitative difference in the behaviour of the Binder cumulant $B$ when sampled at different energy separations. Fig.~\ref{fig:MoreBs} shows the median Binder cumulant (as in Fig.~\ref{fig:B_vs_L}(a) in the main text) for $\omega = 0.25, 0.5, 1.5$.
We include uncertainty esimates in these plots as we have simulated fewer disorder realisations than for $\omega=1$; the errorbars indicate $\sqrt{2/N} / f_{\rm Med}$, where $f_{\rm Med}$ is the value of the probability distribution of $B$ at its median value.

\begin{figure}
    \centering
    \includegraphics[width=\columnwidth]{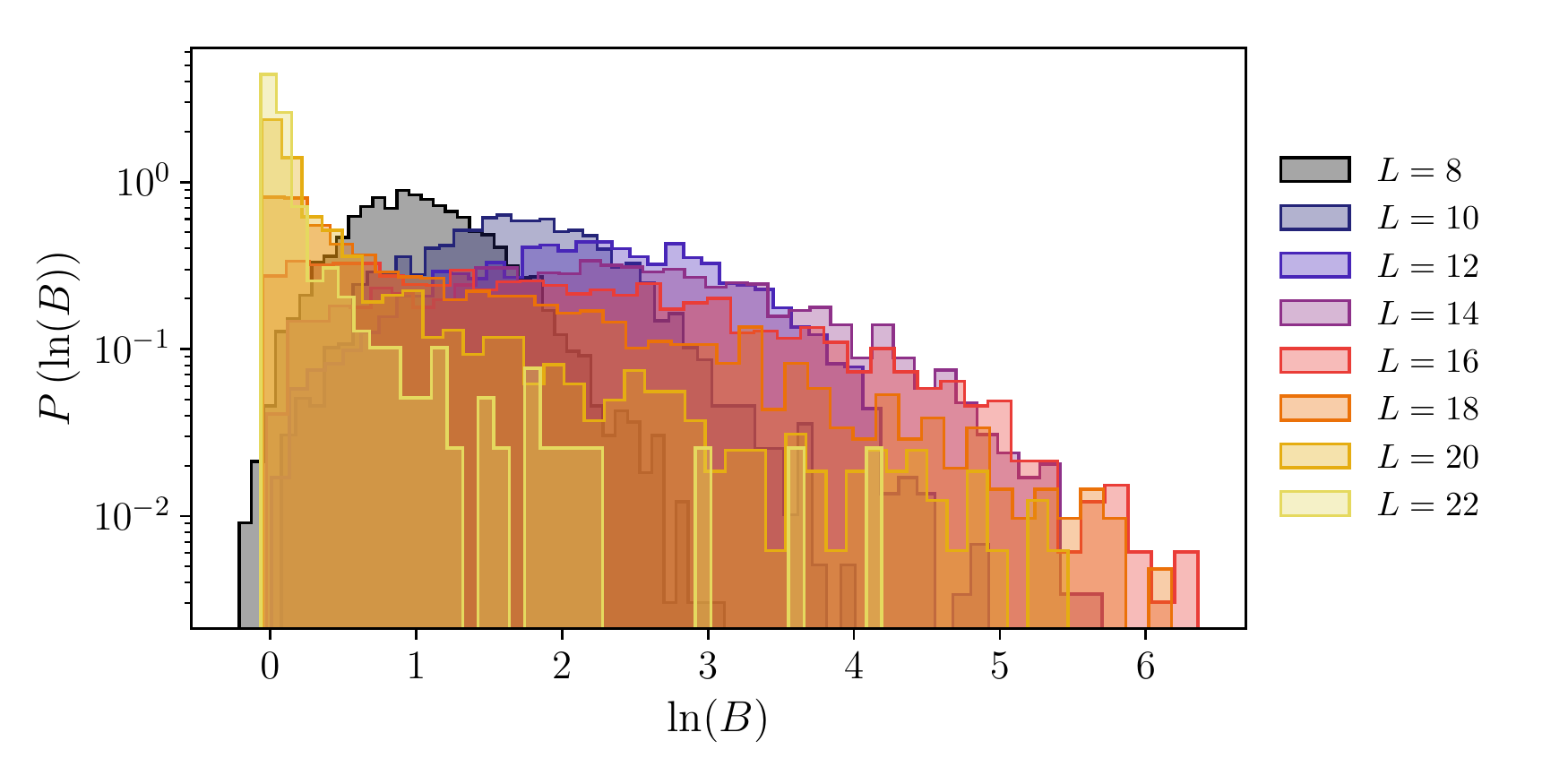}
    \caption{Histograms of the logarithm of the Binder cumulant $\ln B$ across disorder realisations for several system sizes $L$, shown for $\omega=1$ and $W = 1.8$.}
    \label{fig:BHists}
\end{figure}

Fig.~\ref{fig:BHists} shows histograms of $B$ for several system sizes with $W=1.8$, demonstrating how the large values of $B$ near $L_0$ originate from broad distributions.
As $L$ becomes larger than $L_0$ (in this case $L_0(W=1.8) \approx 14$) the distributions become increasingly narrow, and concentrated around the Gaussian value $B=1$.

\section{Predictions of SDRG}

In the SDRG the main objects of interest are the density of ``ergodic" blocks (or ``bubbles"), $\rho$, and the typical length of the decay of the influence of such block on its surrounding,  $\zeta$, or better its inverse $z=\zeta^{-1}$. The equations (from Ref.~\cite{dumitrescu2019kosterlitz}) are
\begin{eqnarray}
    \frac{dz}{d\ell}&=&-c\rho z+...\\
    \frac{d\rho}{d\ell}&=&b\rho(z^{-1}-z_c^{-1})+...\ ,
\end{eqnarray}
these equations flow as the scale $\ell$ grows from large $z$ to small $z$. The scale $\ell$ is interpreted as the logarithm of a length and $\sim \ln L$ ($L$ being the system size) at the end of the evolution. The flow of $\rho$ is downward for $z>z_c$ and upward for $z<z_c$. So if $\rho=0$ is reached before $z=z_c=\ln 2$ the system ends in an MBL phase. Otherwise the flow inverts and what initially looked like an MBL phase turns into an ETH phase. It is tempting to interpret the system size at which this happens as our $L_0(W)$ introduced in the main text. To do this we can solve the equations as
\begin{equation}
    \ell=\int_{z,\rho}^{z_0,\rho_0}\frac{dz}{cz\rho}.
\end{equation}
and
\begin{equation}
    \rho=\gamma/z+\gamma\ln z/z_c+C,
\end{equation}
where $\gamma=b/c$ and $C$ is the integration constant which defines the trajectory. Close to $W_c$, $C$ is regular $C\simeq C_c+\alpha(W_c-W)+...$ (with $\alpha>0$ and $C_c$ immaterial), and so $\rho\simeq (z-z_c)^2+\alpha \Delta W$. At the same time, for the purpose of looking for small $\Delta W$, $z_0,\rho_0$ can be sent to $\infty$ as the integral converges on the upper end:
\begin{equation}
    \ell\simeq\int_{z_c}^\infty\frac{dz}{\alpha \Delta W z+(z-z_c)^2/(2 z_c^2)}.
\end{equation}
As the integral diverges for $\Delta W\to 0$, expanding for small $\Delta W$ gives
\begin{equation}
    \ell\simeq\pi\frac{\ln 2}{2\alpha}\Delta W^{-1/2}+...\ .
\end{equation}
Again, upon interpreting $\ell$ as the logarithm of the system size we have the prediction quoted in the text
\begin{equation}
    L_0\sim {\rm e}^{c \Delta W^{-1/2}},
\end{equation}
for some constant $c$, at leading order. It is worth noting that exactly at the turning point we have
\begin{eqnarray}
    \rho_{L_0}\sim \Delta W.
\end{eqnarray}
It would be interesting to connect $\rho$ with $B$ to make a prediction of how $B$ should diverge at the transition (a zero density of ergodic bubbles would imply a diverging $B$ as the matrix elements would follow a power-law distribution).

\end{document}